\documentclass[superscriptaddress, reprint, amsmath, amssymb, aps, pra, floatfix]{revtex4-2}
\usepackage[utf8]{inputenc}

\usepackage[colorlinks=true]{hyperref}
\usepackage{graphicx}
\usepackage{dcolumn}
\usepackage{bm}
\usepackage{orcidlink}
\usepackage{physics}
\usepackage{color}
\usepackage{amsthm}
\definecolor{pink}{rgb}{1,0,0.9}

\definecolor{purple}{rgb}{0.63,0,1}

\definecolor{pink}{rgb}{1,0,0.9}
\newcommand{\change}[1]{{#1}}   

\begin{document}
\title{Single-layer tensor network approach for three-dimensional quantum systems}
\author{I.V. Lukin\orcidlink{0000-0002-8133-2829}}
\email{illya.lukin11@gmail.com}
\affiliation{Karazin Kharkiv National University, Svobody Square 4, 61022 Kharkiv, Ukraine}
\affiliation{Akhiezer Institute for Theoretical Physics, NSC KIPT, Akademichna 1, 61108 Kharkiv, Ukraine}

\author{A.G. Sotnikov\orcidlink{0000-0002-3632-4790}}
\email{a\_sotnikov@kipt.kharkov.ua}
\affiliation{Karazin Kharkiv National University, Svobody Square 4, 61022 Kharkiv, Ukraine}
\affiliation{Akhiezer Institute for Theoretical Physics, NSC KIPT, Akademichna 1, 61108 Kharkiv, Ukraine}
\date{\today}

\begin{abstract}
    Calculation of observables with three-dimensional projected entangled pair states is generally hard, as it requires a contraction of complex multi-layer tensor networks. We utilize the multi-layer structure of these tensor networks to largely simplify the contraction. The proposed approach involves the usage of the layer structure both to simplify the search for the boundary projected entangled pair states and the single-layer mapping of the final corner transfer matrix renormalization group contraction. We benchmark our results on the cubic lattice Heisenberg model, reaching the bond dimension $D=7$, and find a good agreement with the previous results.  
\end{abstract}

\maketitle

\section{Introduction}

Three-dimensional (3d) lattice systems are a fact of our everyday experience, since they are the natural building blocks of quantum materials. From the theoretical point of view, the 3d physics can be described to some extent by the mean-field approaches. However, these systems are still capable of hosting exotic phases and phenomena, such as the 3d topological orders \cite{Canals_1998, Savary_2017}, topological \cite{hasan2010colloquium} and hinge insulators \cite{benalcazar2017electric}.
Furthermore, access and control over all three spatial degrees of freedom in cold gases of neutral atoms plays a crucial role in realizations of quantum simulators with these systems~\cite{Gross2017}. The neutral atoms subjected to periodic potentials of optical lattices enable direct access to not only colossuses of condensed matter physics, such as the spin-1/2 Hubbard and Heisenberg models, but also to more exotic multiflavor phenomena thermodynamically more accessible in three spatial dimensions~\cite{Taie2022}. Within cold-atom realizations one can also directly study the effects of the dimensional crossover on the topological Mott insulators~\cite{Scheurer2015} and exotic magnetic orders~\cite{unukovych2024anisotropy}.

Therefore, 3d quantum lattice systems become an excellent platform for the development and application of the beyond-mean-field approaches, which are capable of capturing the non-trivial quantum entanglement phenomena and, in particular, the topological order. Tensor networks (TN) \cite{2020_cirac, 2022_Okunishi, Orus2019review, 2023_Banuls} are one of the most powerful and accurate approaches in this direction. Conceptually, it is possible to generalize the TN wave functions to 3d systems, but the numerical cost limitations become severe in practice.

Here, let us focus on applications of tensor networks to quantum systems in the zero-temperature limit. These can be employed either to evaluate the zero-temperature partition function of the quantum system using the tensor renormalization group \cite{Levin_2007, gu2009tensor, 2022_Meurice} (transfer matrix approach) or as a variational ansatz for the ground-state wave function. In the latter type of approaches the ground-state wave function is approximated as a network of tensors. If the Hamiltonian is local, the approximation is justified by the area law of entanglement \cite{Hastings_2007, Eisert_2010}.  For the one-dimensional (1d) system, one usually constructs the matrix-product state (MPS) architecture~\cite{SCHOLLWOCK2011, 2020_cirac} of the tensor network as a variational ansatz and uses the density-matrix renormalization group approach~\cite{white_1992} as a ground-state search algorithm.


For 2d systems, the TN methods, which are based on the projected entangled pair states (PEPS) wave functions \cite{verstraete2004renormalization, Verstraete_2004, nishio2004tensor} (as well as on multiscale entanglement renormalization ansatz \cite{2008_Cincio} or tree tensor networks \cite{Tagliozzo_2009}), allow for very accurate results for a large number of strongly correlated lattice problems, including the Hubbard \cite{Corboz_2016, Corboz_2017} and Heisenberg models on various lattices \cite{2017_xiang2, Corboz_2011, Corboz_2012, Corboz_2013, Bauer_2012, Mila_2012}.  The natural question is whether we can extend all these achievements to the 3d systems. By noticing that the entanglement monogamy \cite{CKW,Osborne_2006} usually simplifies the 3d problems, we expect that the entanglement between the neighbors of the certain range will be smaller in 3d due to a larger number of these neighbors \cite{Osborne_2006,Osterloh_2015}.  This fact allows us to restrict to rather low bond dimensions $D$ of corresponding tensors to achieve accurate results.  Still, we are aiming at a scheme that enables analysis with moderate values of $D$ (in particular, $D\lesssim8$) and affordable speed of calculations, as well as hosting a possibility of results extrapolation to the infinite-$D$ limit. 

The previous studies on 3d tensor networks utilized different strategies. In particular, in Refs.~\cite{2019_jahromi, 2020_jahromi, 2021_jahromi, 2016_rakov} the authors employed the simple update scheme \cite{Jiang_2008} to both optimize the PEPS for infinite systems (iPEPS) and to calculate the observables. This approach is efficient for the gapped systems. In Ref.~\cite{2013_latorre} a type of the tensor renormalization group was applied for the tensor network contraction. In Ref.~\cite{2021_tepaske} a special isometric tensor network was employed, which allows for the exact calculation of observables. In the study~\cite{Magnifico_2021} the tree tensor networks were applied to 3d lattice gauge theory on a finite lattice.  Within our study, we follow the approach given in Refs.~\cite{2021_vlaar, 2023_vlaar, 2023_vlaar2}, which uses iPEPS wave function on 
3d cubic lattice and employs the boundary iPEPS and corner transfer matrix renormalization group (CTMRG) \cite{Nishino_1996, Nishino_1997, Orus_2009} for the calculation of observables. 

In this paper,  we follow the simple update methodology to optimize the 3d iPEPS on the cubic lattice, whilst proposing and testing \emph{a single-layer computational procedure} (based on ideas from Refs. \cite{2017_xiang, 2019_Hagshenas}) for calculating observables with it. 
The approach extensively utilizes the layered structure of the tensor network to reduce the scaling with the bond dimension at the expense of the enlargement of the elementary unit cell. The beneficial property of this method is its $D^{12} - D^{13}$ computational scaling (depending on specific details), which is comparable to the cost of the typical calculations with the 2d iPEPS approach.  Let us emphasize that the latter corresponds to the cost of computation of observables and not the cost of the iPEPS optimization, which is typically smaller. Hence, this computation should be performed only once during the calculation (at least, if one follows the simple update strategy in the iPEPS optimization). Our approach is partially connected with a recent study~\cite{Yang_2023} with the single-layer TN contraction applied to a classical 3d statistical mechanics problem.

\section{Algorithm description}

\subsection{iPEPS and its symmetries}

\begin{figure}
    \includegraphics[width= \linewidth]{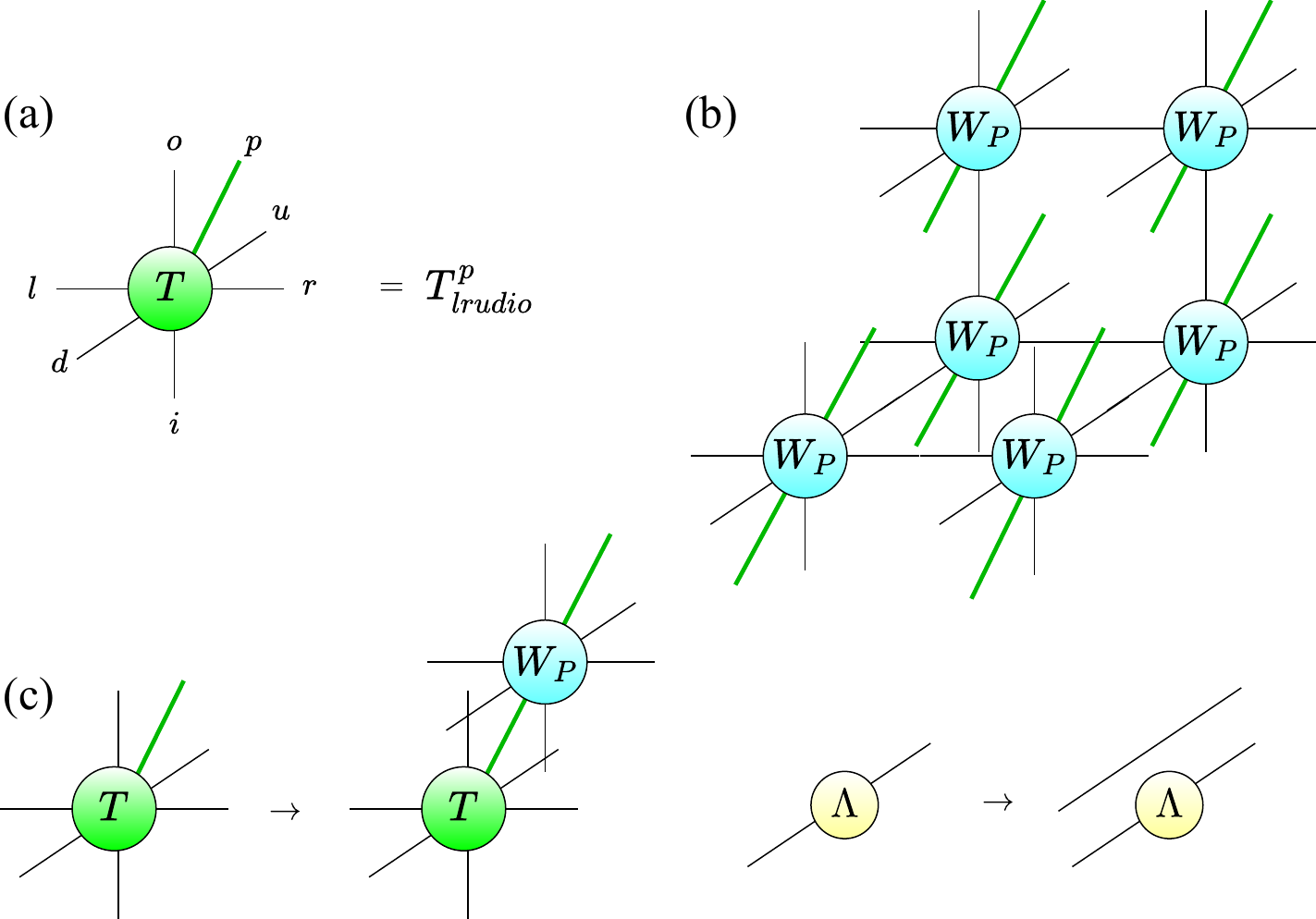} 
     \caption{\label{fig: 3d_PEPO}%
       (a) The individual tensor $T$ of the iPEPS wave function. (b) The graphical illustration of the 3d PEPO, which consists of individual tensors $W_{P}$. The tensors are chosen to approximate the operator $\exp(-H dt)$. (c) The application of PEPO to individual tensors $T$ of the 3d iPEPS. The application leads to an enlarged bond dimension $D$, which is the product of the bond dimensions of the iPEPS and PEPO. The application of PEPO to iPEPS including the positive matrices \change{$\Lambda$} results in the tensor network with the enlarged number of bonds, where the matrices \change{$\Lambda$} remain the same, while the additional (side) lines include the Kronecker deltas.
       }
\end{figure}

Within the current study, we focus on the ground-state properties of 3d quantum many-body systems on the simple cubic lattice. To model the wave function as a tensor network of the iPEPS type~\cite{2021_bruognolo}, we place the rank-7 tensors $T^{p}_{lrupio}$ on all the sites of the lattice and positive matrices $\lambda_{lr}$ on all the links of the lattice, as shown in Fig.~\ref{fig: 3d_PEPO}(a). The index $p$ corresponds to the physical on-site Hilbert space, while other $l,r,u,d,i,o$ correspond to the virtual indices (left, right, up, down, in, and out, respectively). To obtain the iPEPS wave function, all the virtual indices are contracted according to the spatial geometry of the lattice. 

Below, we make additional assumptions on the tensors~$T$. First, we assume that the tensors are the same for all sites of the lattice. Hence, the wave function is translationally invariant.  The assumption of translational invariance can be broken in some models with the corresponding generalizations of the construction introduced in this study. Second, we assume that the tensors obey a certain reflection symmetry, e.g., they remain invariant upon the interchange of the left and right, as well as other two pairs of indices, $T^{p}_{lrudio} = T^{p}_{rludio} = T^{p}_{lrduio} = T^{p}_{lrudoi}$. These reflection symmetries guarantee the hermiticity of the transfer matrix of the iPEPS wave function.

\subsection{iPEPS optimization}
We perform the iPEPS optimization using the simple update method with a projected entangled pair operator (PEPO) approximating the evolution (time-stepper) operator $\exp(-H dt)$, where $H$ is the system Hamiltonian and $dt$ characterizes the time step. We do not employ the most popular approach based on the Trotter gate application, since the gate-based scheme requires the enlarged unit cell and breaks rotational and reflection symmetries (at least, during the optimization).  In contrast, the PEPO-based evolution can be used with the $1\times1\times1$ elementary unit cell and can be followed by the explicit tensor symmetrizations after every PEPO application. This allows us to obtain the iPEPS tensors with implicit reflection symmetries and minimal unit cell, which largely simplifies the calculations of observables.

\begin{figure}
    \includegraphics[width= \linewidth]{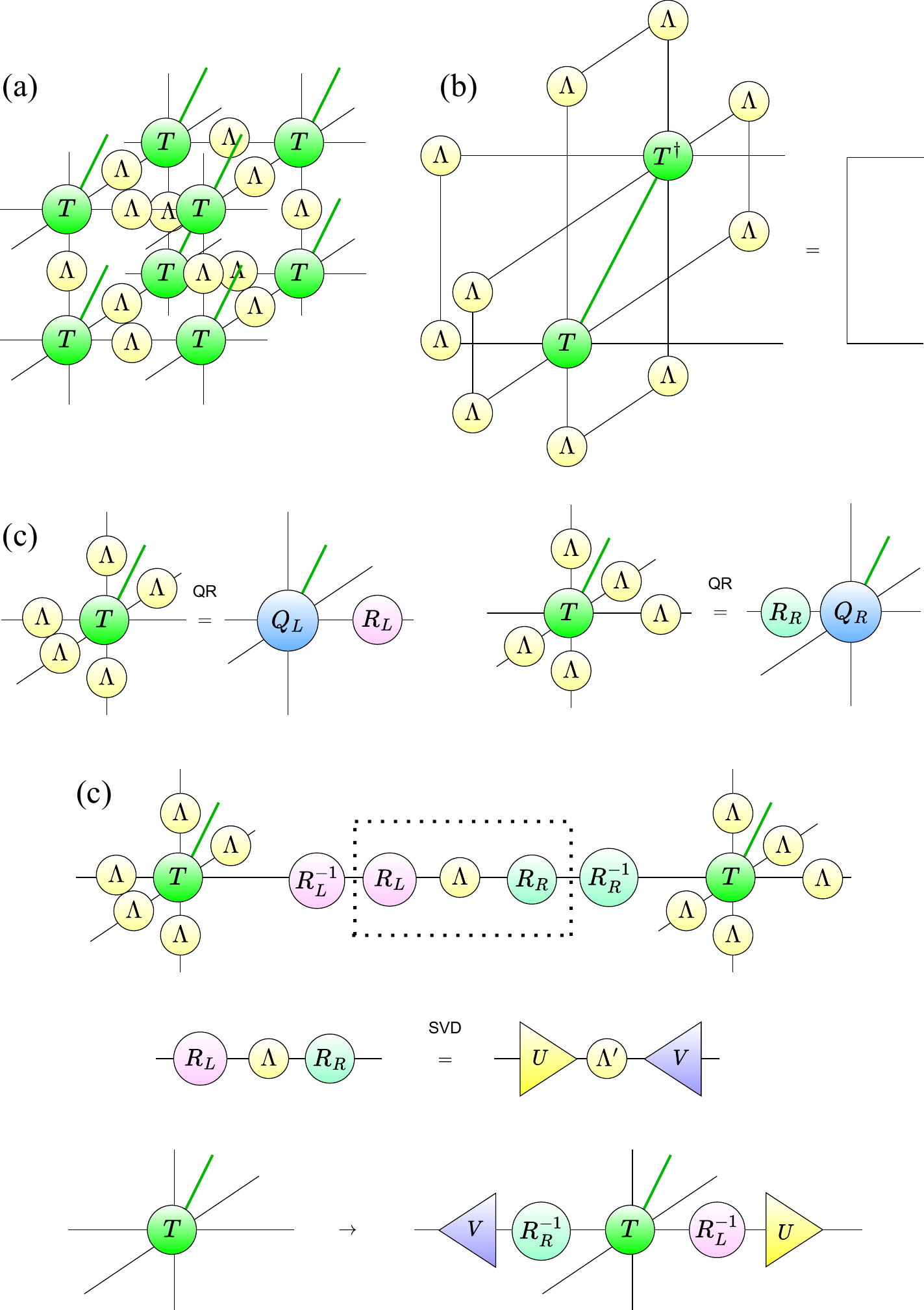} 
     \caption{\label{fig: Canonical_form}%
       (a) The 3d iPEPS wave function, consisting of the bulk tensors $T$ and positive bond matrices \change{$\Lambda$}. (b) The condition of the superorthogonal canonical form, where the right-hand side of the graphical equation is the identity matrix. (c) The iterative scheme of the iPEPS superorthogonalization \cite{2015_phien, 2017_ran}. The fixed point of the iteration is the iPEPS tensors in the canonical form, which can be used for the bonds truncation according to the weights in \change{$\Lambda$}.  }
\end{figure}

The optimization in terms of PEPO runs in several steps. First, we initialize the iPEPS wave function. We use the tensors with $D=1$ as the initial iPEPS, since this initialization speeds up the convergence according to our observations. Second, we determine the PEPO, which approximates the operator $\exp(-H dt)$. This can be done, e.g., by using the $W^{I}$ approach or other cluster-based methods \cite{2015_zaletel, Vanhecke_2021, vanhecke2021simulating} \change{(see also Appendix ~\ref{app:A} for more details)}. After that, we repeatedly apply the PEPO to the iPEPS wave function, as \change{illustrated in Figs.~\ref{fig: 3d_PEPO}(b), \ref{fig: 3d_PEPO}(c), and \ref{fig: Canonical_form}(a)}. The bond dimension $D$ grows with every PEPO application. Hence, after several first applications one needs to start truncating the iPEPS bond dimension back to some fixed target bond dimension $D$. The truncation can be performed by using the superorthogonal canonical form~\cite{2015_phien, 2017_ran,2023_tindall, Ran_2012}, which is defined by the condition shown in Fig.~\ref{fig: Canonical_form}(b). 
\change{This condition should hold for all tensors and bonds of the iPEPS wave function and it is a natural generalization of the canonical form of matrix product states.} 

To truncate the iPEPS bond dimension, we iteratively transform it to the canonical form, as in Fig.~\ref{fig: Canonical_form}(c), and then truncate the bonds according to the magnitudes of diagonal elements of \change{$\Lambda$} (we truncate the smallest diagonal elements). Note that in the superorthogonal canonical form, the matrices \change{$\Lambda$} are diagonal and positive. 
\change{The transformation into the canonical form can be performed as shown in Fig.~\ref{fig: Canonical_form}(c): First, one can take one particular bond of the iPEPS wave function with the two tensors~$T$ on this bond and absorb all matrices~$\Lambda$ on the adjacent bonds into these tensors. Next, one can decompose these tensors by means of the QR decomposition, as we also show in Fig.~\ref{fig: Canonical_form}(c). The resulting matrices $R_{L}$ and $R_{R}$ can be viewed as obstructions to the conditions in Fig.~\ref{fig: Canonical_form}(b). To get rid of these matrices, one can simultaneously insert the matrices $R_{L}$ and $R_{R}$ and their inverses on the bonds. Inverses can be then absorbed into the tensors~$T$, while the matrices $R_{L}$ and $R_{R}$ are absorbed into the bond matrix~$\Lambda$. The resulting bond matrix (which is no longer diagonal) is now decomposed with the singular value decomposition (SVD), $R_L \Lambda R_R = U \Lambda' V$, and the corresponding SVD unitary matrices $U$ and $V$ are also absorbed into the tensors~$T$, while the SVD spectra form the new positive and diagonal matrices~$\Lambda'$. The iterations can be repeated until convergence of the bond matrices~$\Lambda$.  }

By repeating these PEPO applications, canonical form iterations, and truncations, we converge the iPEPS wave function to the true ground state. The convergence can be monitored either with the matrices \change{$\Lambda$} or observables computed with the Simple Update scheme. Generally, we observe that the converged tensors are symmetric under reflections up to a very high accuracy with the norm of the non-symmetric part being of the order $10^{-12}$.

\subsection{Boundary iPEPS: Simple update scheme}

\begin{figure}
    \includegraphics[width= \linewidth]{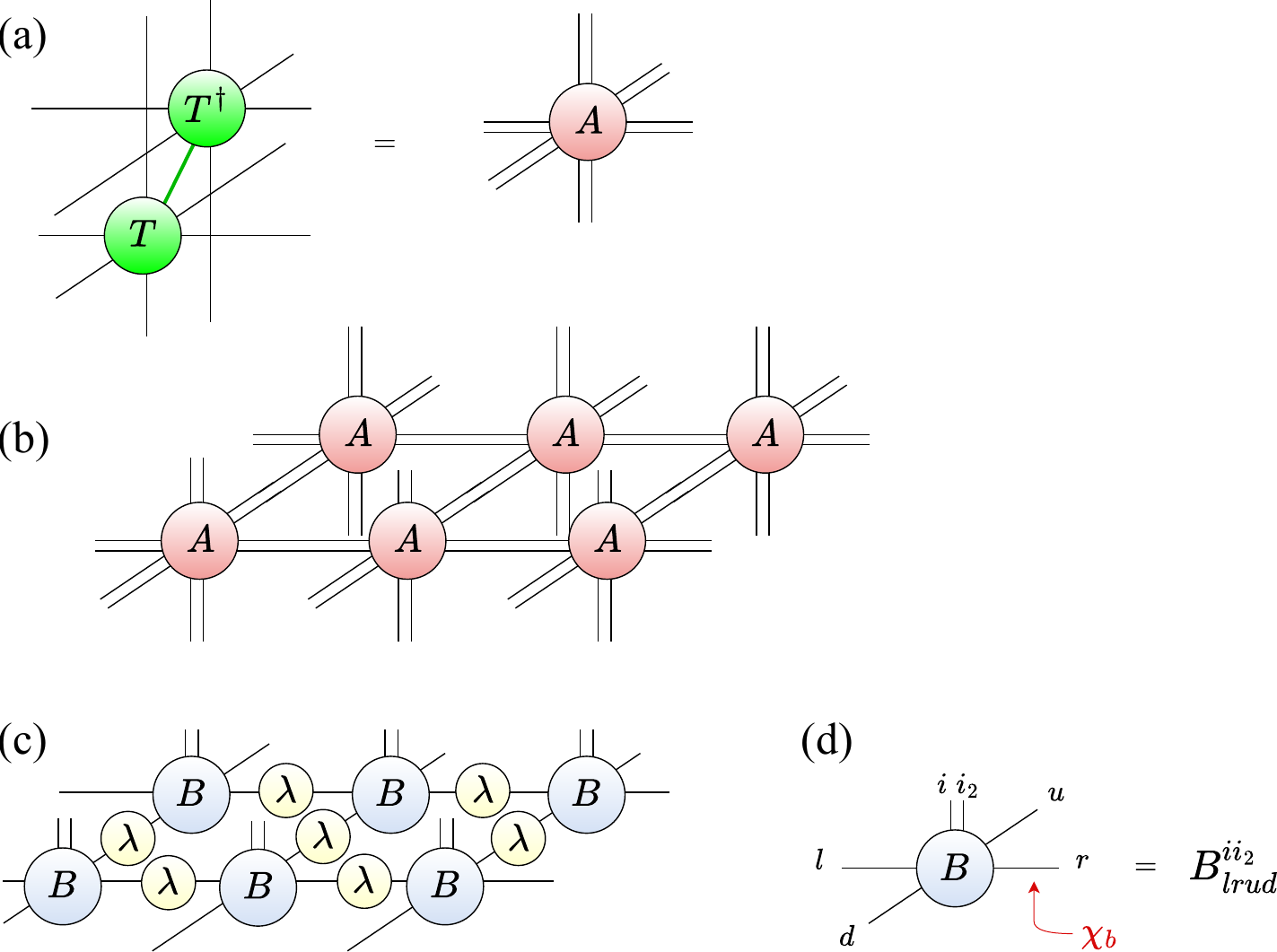} 
     \caption{\label{fig: FigurebPEPS}%
       (a) Definition of the double-layer tensor $A$ and (b) infinite 3d tensor network composed of these tensors (the transfer matrix $M$ consists of all tensors $A$ on the horizontal plane); (c) The left boundary PEPS (bPEPS) consists of identical tensors $B$ and positive matrices $\lambda$ placed on its virtual links. (d) Each tensor $B$ has the bond dimension~$\chi_{b}$ for its four virtual indices.} 
\end{figure}

Upon completing the iPEPS optimization, either by applying the above PEPO scheme or Trotterized gates, it is necessary to compute observables with the obtained wave function. To this end, we need to find an efficient method to compute the wave function norm $\langle \Psi| \Psi \rangle$ and the local operator insertions $\langle \Psi| \hat{O} |\Psi\rangle$, where $\hat{O}$ is the local operator.  The wave function norm can be cast in the form of contraction of the two infinite 3d tensor networks, which consists of the double-layer tensors $A = T^{p}_{lrudio}\bar{T}^{p}_{l_{2}r_{2}u_{2}d_{2}i_{2}o_{2}}$, where the pairs of indices, e.g., $l$ and $l_{2}$, are combined into a double-layer index, as shown in \change{Fig.~\ref{fig: FigurebPEPS}(a)}. This double-layer index has the dimension $D^{2}$ (note that we absorb the bond matrices into the tensors~$T$).

To contract this network, we employ a transfer-matrix method: first, we write the tensor network contraction as $\langle \Psi| \Psi \rangle = \Tr(M^{N})$, where $M$ is the planar transfer matrix, which consists of the double-layer tensors~$A$ in the infinite plane with all in-plane indices contracted and all indices perpendicular to the plane considered to be the matrix indices, \change{as shown in Fig.~\ref{fig: FigurebPEPS}(b)}.  We compute the left and right leading eigenvectors of the transfer matrix $M$ as $\langle l| M = \lambda \langle l|$ and $M|r\rangle = \lambda|r\rangle$.  In the limit of the infinite lattice, these are sufficient to compute the wave function norm and all the \change{nearest-neighbor} operator averages.  For the models studied here, the transfer matrix~$M$ is generally real and symmetric, thus the left and right leading eigenvectors coincide. Still, all the schemes discussed below hold for the non-symmetric transfer matrix $M$ with the condition that all calculations must be repeated for both $\langle l|$ and $|r\rangle$. 

To determine the leading eigenvectors of $M$, we follow the suggestion from Ref.~\cite{2021_vlaar}, i.e., approximate these leading eigenvectors as the two-dimensional iPEPS of the bond dimension $\chi_{b}$ with the role of physical index replaced by the combined index of double-layer tensor $A$ in the transverse direction to the transfer matrix plane, \change{as depicted in Fig.~\ref{fig: FigurebPEPS}(c)}. For example, if the transfer-matrix layers are in the $xy$ plane, then the indices $i,o$ are transverse, and physical indices of the left iPEPS tensors are $i,i_{2}$, while the physical indices of right boundary iPEPS are $o, o_{2}$. In total, the left boundary iPEPS consists of the rank-6 tensors $B^{i i_{2}}_{lrud}$, with $i$ and $ i_{2}$ of the dimension $D$ (from the $T$ tensors of 3d iPEPS), while $l,r,u,d$ are the auxiliary indices of the dimension $\chi_{b}$, as shown in Fig.~\ref{fig: FigurebPEPS}(d). Note that in the case of the enlarged unit cell of the bulk 3d iPEPS, the boundary iPEPS tensors $B$ must be also taken with the enlarged unit cell, which is the projection of the 3d unit cell on the boundary plane. 

\change{Our next step} is to determine the left boundary iPEPS tensors~$B$. In this section, we follow another suggestion from Ref. \cite{2021_vlaar} and adapt the simple update scheme to find $B$. Still, our implementation of \change{this update} is different from Ref.~\cite{2021_vlaar}. We propose to use explicitly the double-layer structure of the transfer matrix tensors $A$ to reduce the scaling of the simple update calculations to $\chi_{b}^{5} D^{7}$. 

\begin{figure}
    \includegraphics[width= \linewidth]{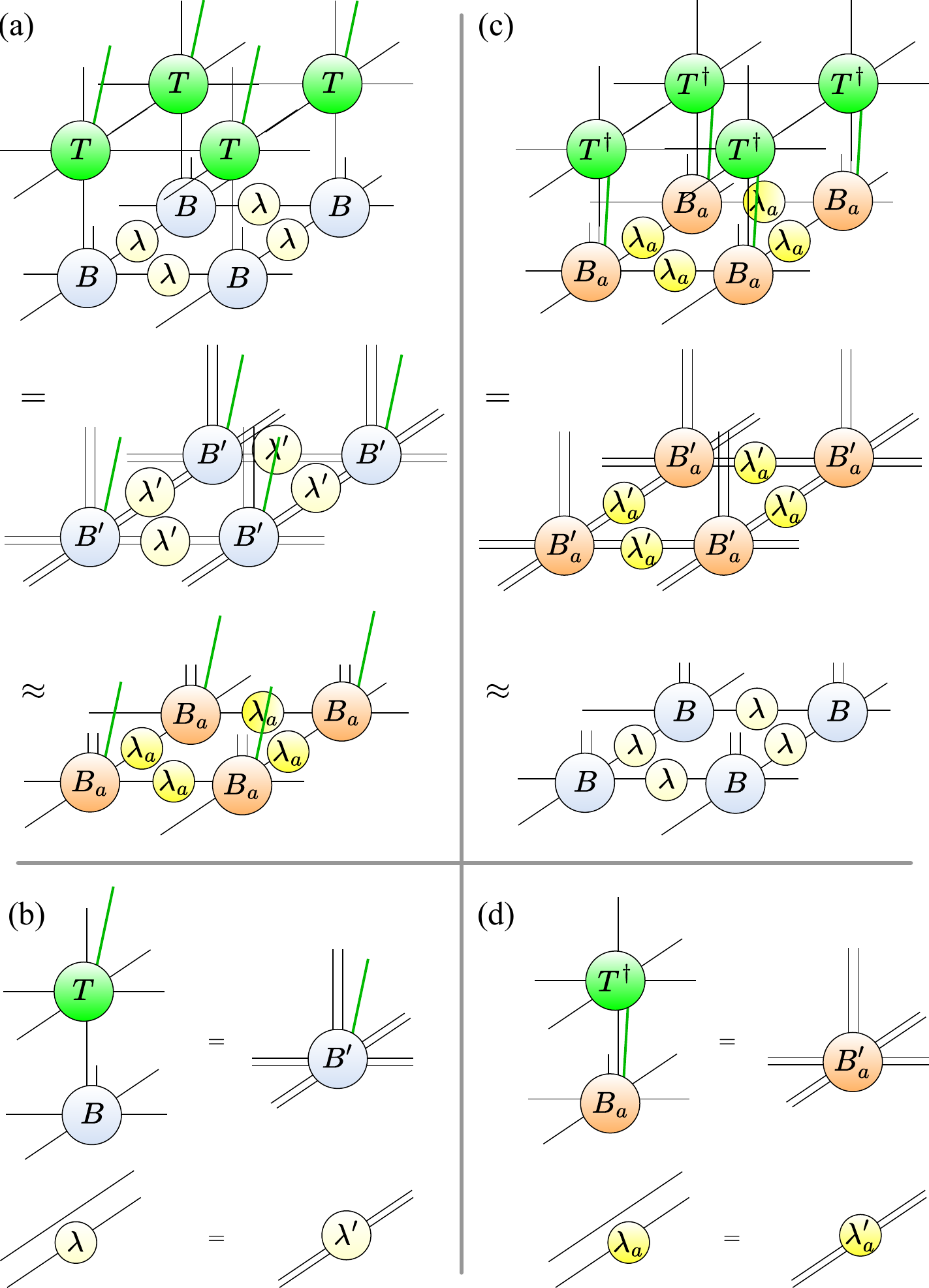} 
     \caption{\label{fig:FigureUpdate}%
        (a) and (b) Transfer matrix~$M$ can be applied to the bPEPS in a layered fashion. First, the tensors $T$ are absorbed into the bPEPS tensors $B$ to obtain the new enlarged tensor $B'$.  Next, the bond dimension of the tensor $B'$ is truncated back to the original dimension $\chi_{b}$, which results in the auxiliary (intermediate) tensor $B_{a}$.  (c) and (d) Similar procedure with the second layer composed of $T^{\dagger}$ tensors: We absorb $T^{\dagger}$ into $B_{a}$ and then truncate the enlarged tensors~$B'_{a}$ back to the original dimension $\chi_{b}$. This yields the original bPEPS tensors $B$. The iterations are repeated till convergence of the tensor~$B$ and the corresponding matrix~$\lambda$.}
\end{figure}

We propose to apply the transfer matrix $M$ to the bPEPS layer by layer, i.e., first, we take the layer consisting of only $T$, as shown in Fig.~\ref{fig:FigureUpdate}(a). The tensors~$T$ are then absorbed into the bPEPS. This step has the computational scaling  $\chi_{b}^{4} D^{7}$ and memory scaling $\chi_{b}^{4} D^{6}$. As a result, we obtain the new enlarged tensors~$B'$, with virtual indices of the dimension $\chi_{b} D$. This new enlarged tensor is necessary to truncate back to the original dimension $\chi_{b}$. This can be realized either by employing the simple update canonical form or with some version of the full update scheme (intermediate variants, e.g., the neighborhood tensor update or cluster updates are also possible \cite{dziarmaga2021time, lubasch2014unifying, 2011_cluster, LoopUpdate2020}).

After the truncation, we obtain a new auxiliary tensor~$B_{a}$ (it is auxiliary, since it does not appear in calculations of averages; it is necessary only as an intermediate step in the bPEPS optimization). We also introduce the auxiliary positive bond matrix $\lambda_{a}$ related to the tensor~$B_a$. Next, let us apply the layer of $T^{\dagger}$ tensors to the auxiliary bPEPS of $B_{a}$, as shown in Fig.~\ref{fig:FigureUpdate}(c). Similarly to the above-introduced scheme, we first absorb the tensors $T^{\dagger}$ into $B_{a}$ and then truncate them to the original bond dimension $\chi_{b}$. As a result, we obtain the original bPEPS tensor $B$. This update iteration is repeated until the convergence of $B$. Note that if the original bulk tensors $T$ had a larger unit cell, then the loop may include a larger number of the layer applications.

To complete the description of the bPEPS update, let us add a few words about the truncation procedure. In case one employs the simple update approach, this is performed in the same way as described above in the 3d optimization scheme.

\subsection{Boundary CTMRG: Three-layer approach}

\begin{figure}
    \includegraphics[width= \linewidth]{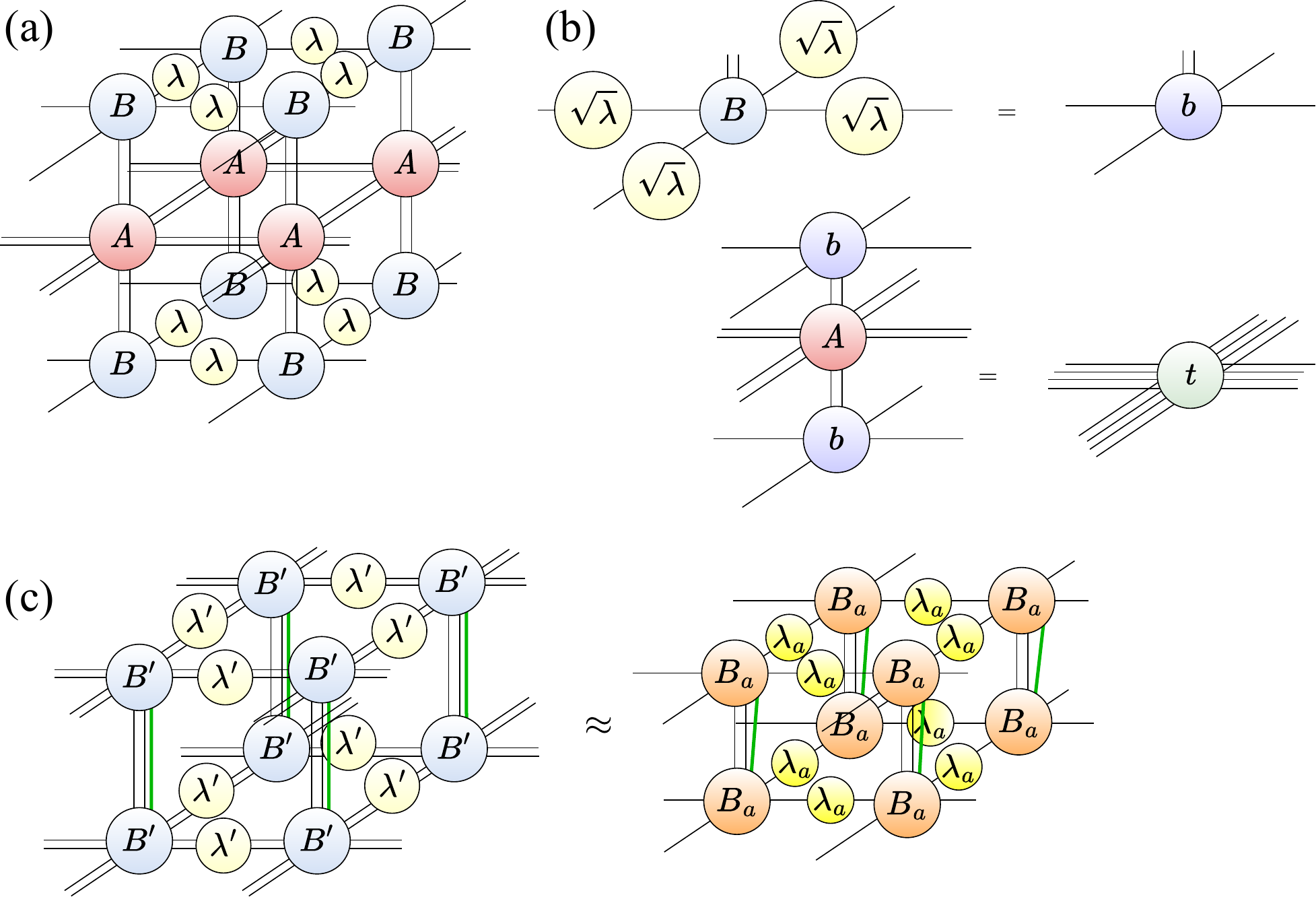} 
     \caption{\label{fig:FigureCTMRG_1layer}%
       (a) The infinite 2d tensor network necessary for the local observables computation. 
       (b) The bond matrices $\lambda$ can be absorbed into the bPEPS tensors $B$, resulting in the new boundary tensors~$b$. The boundary tensors $b$ can sandwich the bulk tensor $A$ into the new single layer tensor $t$ of the bond dimension $D^{2} \chi_{b}^{2}$. 
       (c) Alternatively, one can absorb the tensors $T$ (from $A$) into the two bPEPS layers with the tensors~$B'$. The structure can be further approximated (by the leg dimension truncation back to $\chi_{b}$) to the two-layer tensor network of $B_{a}$. }
\end{figure}

After obtaining the boundary iPEPS, let us turn to the measurements of observables. The observables calculation requires a method to contract an infinite 2d tensor network consisting of three different layers of tensors: A central layer of $A$-tensors and two boundary layers of the boundary iPEPS. This 2d tensor network is shown in Fig.~\ref{fig:FigureCTMRG_1layer}(a). The operator averages can be obtained from this contraction by the additional local operator insertions inside the bulk tensors $A$. This tensor network can be approximately contracted using either CTMRG or boundary MPS methods. To apply CTMRG, we absorb the bond matrices $\lambda$ of bPEPS into the boundary tensors $B$ (resulting in the new tensors $b$) and then contract the tensors $b$ and $A$ into the single-layer tensor $t$, as shown in Fig.~\ref{fig:FigureCTMRG_1layer}(b). The bond dimension of this new tensor $t$ is equal to $D^{2} \chi_{b}^{2}$, which leads to a very costly CTMRG contraction \cite{2021_vlaar}. Hence, it is natural to find a reduced scheme for the 2d tensor network contraction, which will utilize the layered structure of the network in a more efficient manner. 

To find a more efficient scheme, we notice that the tensor network can be cast in the form, which is shown in Fig.~\ref{fig:FigureCTMRG_1layer}(c), where we absorb layers of the tensors $A$ into the boundary iPEPS. Note that this step is performed exactly. The obtained network has only two layers, and, as will be described below, can already be used efficiently to reduce the computational cost. Here, we will use an additional approximate step, and will truncate the dimension of the $B'$ tensors (either with the Simple or Full Update, as is discussed above) to obtain the two-layer tensor network consisting of tensors $B_{a}$ (which have a bond dimension $\chi_{b}$). 

At this stage, we have effectively reduced the tensor network contraction problem to the usual contraction appearing in 2d iPEPS calculations which is generally tractable. Still, the network does have a layered structure. We can use this layered structure, as was proposed in Refs. \cite{2017_xiang, 2017_xiang2, 2019_Hagshenas,2023_lan} to further reduce the computational cost.

\begin{figure}
    \includegraphics[width= \linewidth]{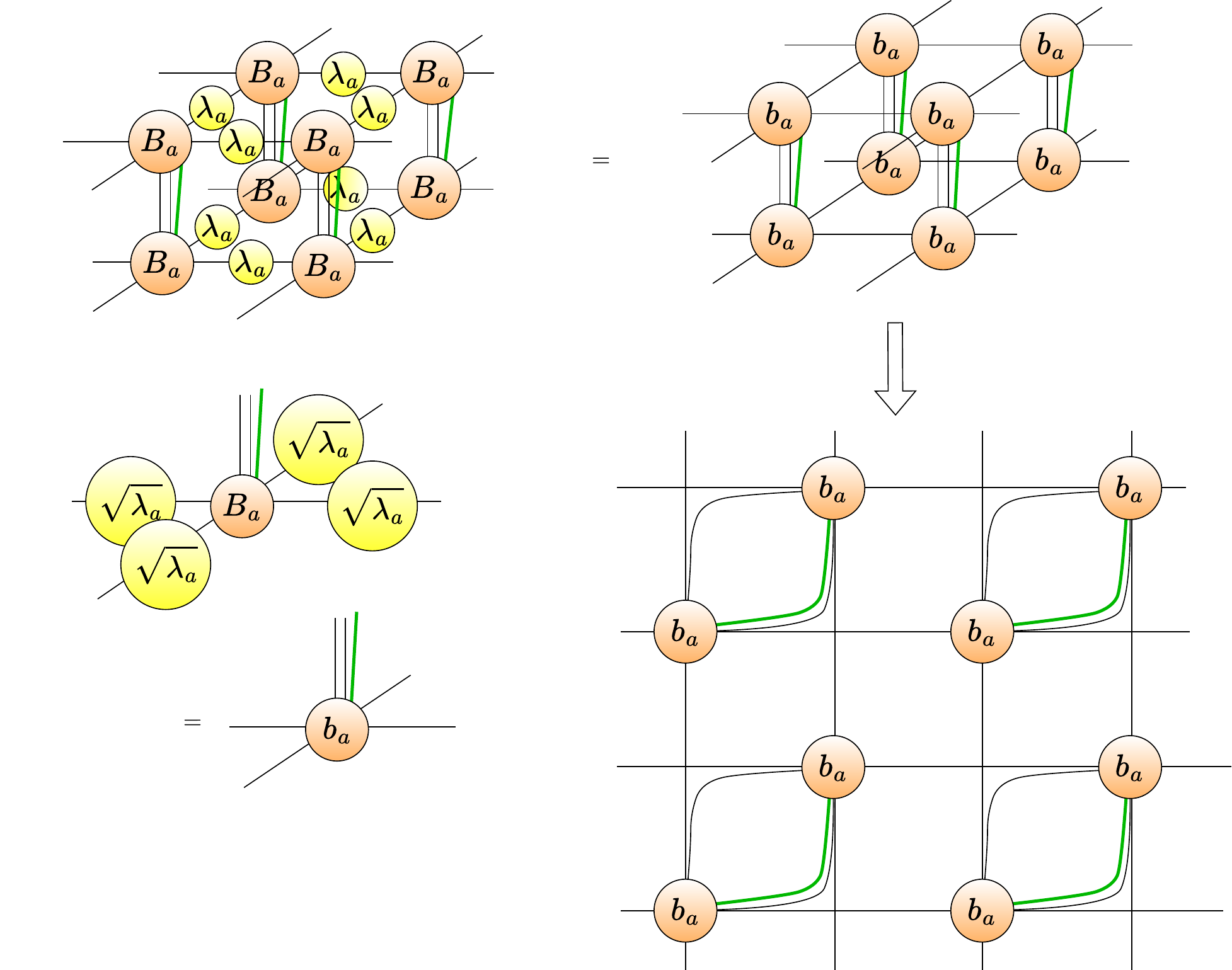} 
     \caption{\label{fig:FigureCTMRG_reduced}%
       An illustrative example of the mapping of the two-layer tensor network with the $1\times 1$ unit cell to the single-layer tensor network with the $2 \times 2$ unit cell. The mapping introduces auxiliary tensors, which consist only of the products of the Kronecker deltas.  }
\end{figure}

A further reduction of the computational cost can be achieved with a mapping of the bilayer tensor network (with the reduced tensors $B_{a}$) into the single-layer tensor network with an enlarged unit cell, as we show in Fig.~\ref{fig:FigureCTMRG_reduced}. This mapping constitutes a two-step procedure, where, first, the bond matrices $\lambda_{a}$ are absorbed into the bulk tensors, and then the bulk tensors are mapped to the $2\times 2$ effective unit cell. This mapping turns the largest effective bond dimension of the network proportional to $\chi_{b} D$, which is asymptotically smaller than $\chi_{b}^{2}$ scaling of the two-layer network. Still, in this case, the benefits of single-layer mapping are not very impressive. 

\begin{figure}
    \includegraphics[width= \linewidth]{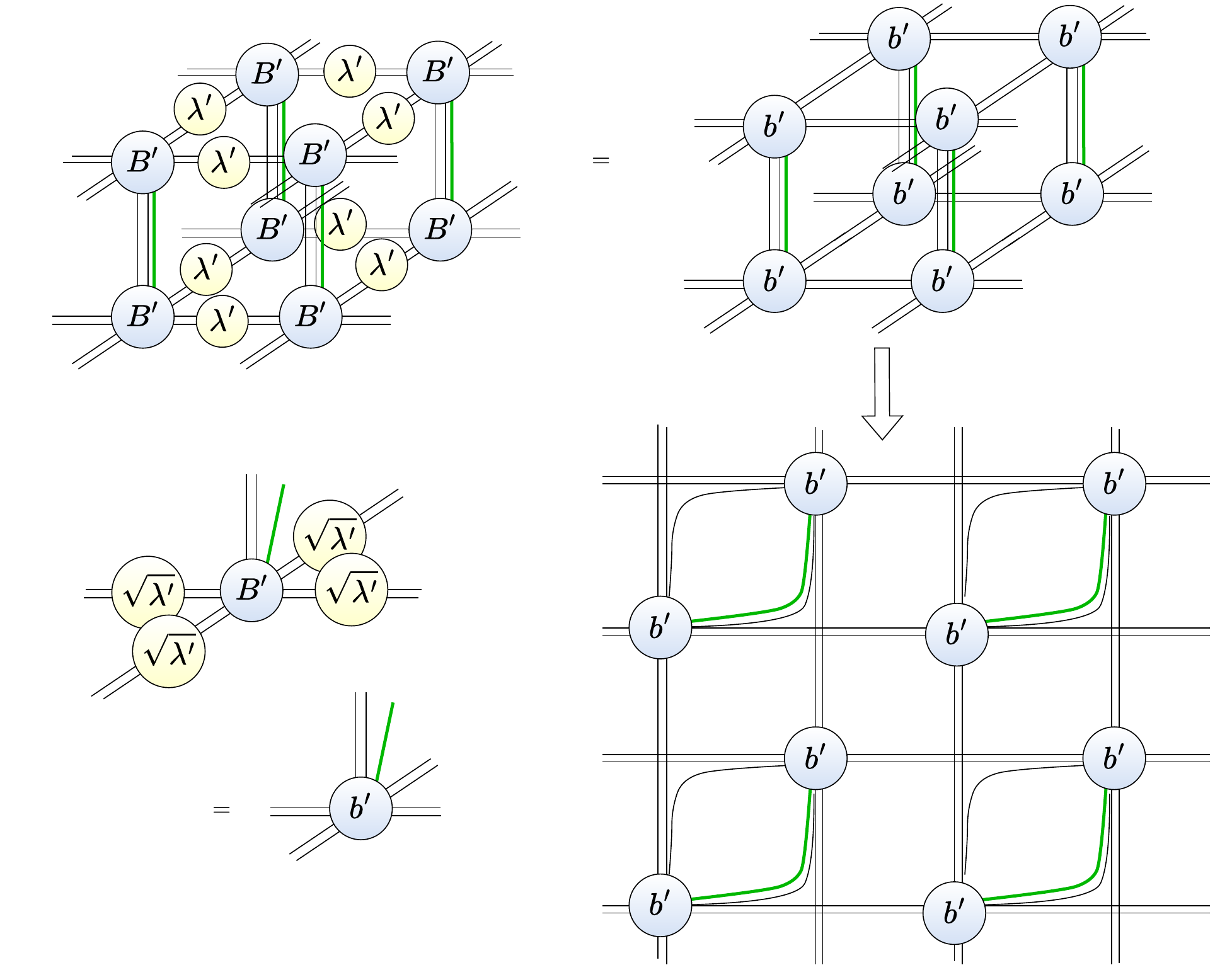} 
     \caption{\label{fig:FigureCTMRG_nonreduced}%
       The mapping of the non-reduced two-layer tensor network with the $1\times 1$ unit cell to the single-layer tensor network with the $2 \times 2$ unit cell. The mapping also introduces auxiliary tensors, which consist only of the products of the Kronecker deltas.}
\end{figure}

We can now turn back to the non-reduced two-layer tensor network, which is shown in Fig.~\ref{fig:FigureCTMRG_1layer}(b) and consists of the tensors~$B'$. For this network, we also have a two-layer structure. Hence, we can employ the same single-layer mapping, as for the reduced tensors. This mapping is shown in Fig.~\ref{fig:FigureCTMRG_nonreduced}. The resulting tensor network has enlarged unit cells, but smaller bond dimensions, with the largest bond dimension scaling as $\chi_{b} D^{2}$ instead of $\chi_{b}^{2} D^{2}$. If the CTMRG bond dimension $\chi$ does not change significantly with the mapping, then we will already obtain a large cost reduction. Note that in this procedure we have not used any truncations of the bond dimensions of the tensors $B'$, thus the resulting single-layer contraction is equivalent to the contraction of the network in Fig.~\ref{fig:FigureCTMRG_1layer}(a). 

\begin{figure}
    \includegraphics[width= \linewidth]{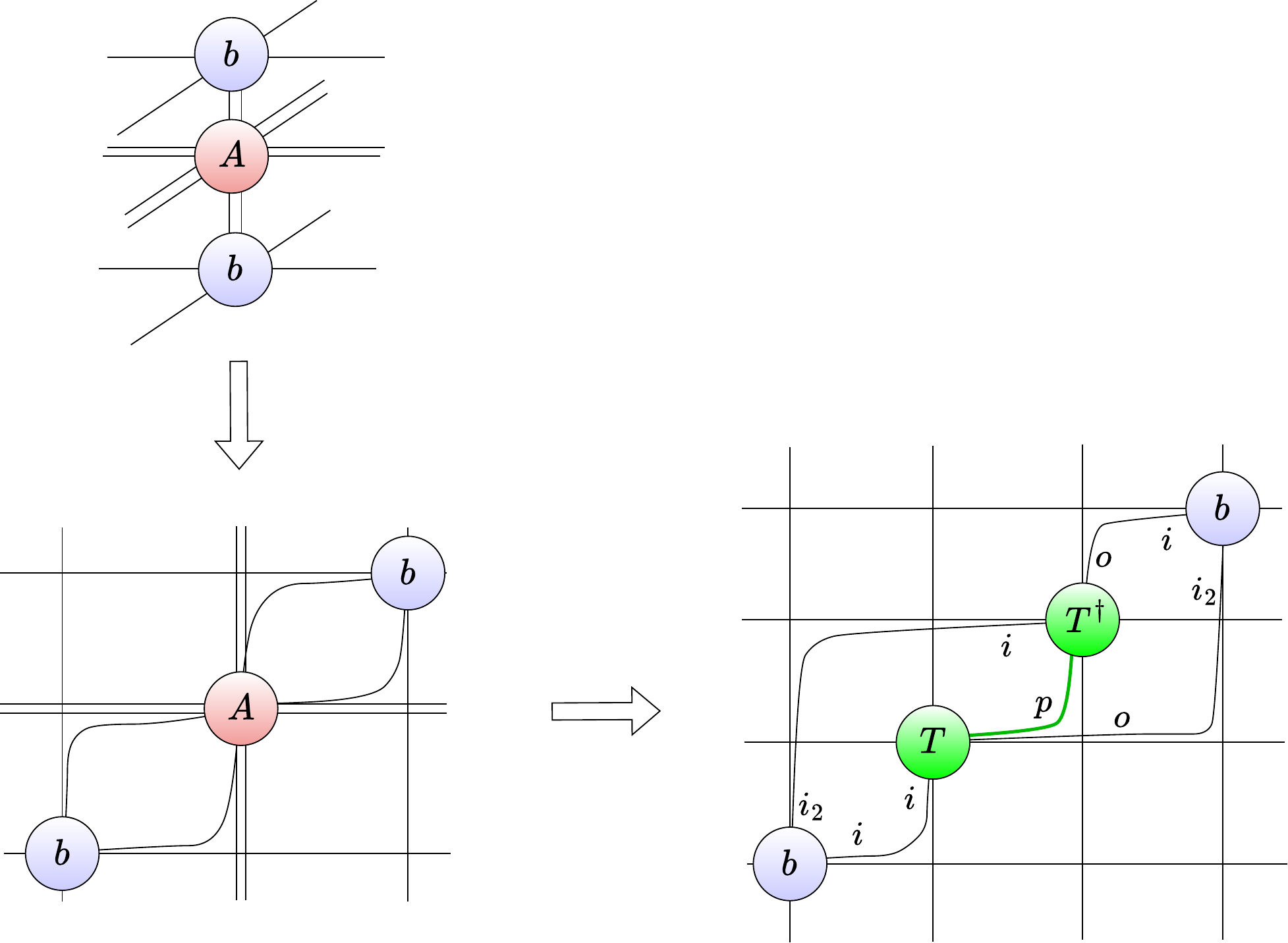} 
     \caption{\label{fig:FigureCTMRG1}%
      The illustration of the mappings of the four-layer tensor network to the single-layer networks with $3\times3$ or $4\times4$ unit cells.  }
\end{figure}

Still, this mapping does not use the full layered structure of the network in Fig.~\ref{fig:FigureCTMRG_1layer}(a), since this network also has a 3-layer and 4-layer structure. It is natural to ask if we can use these 3- or 4-layer forms of the tensor network to devise the single-layer mapping with the enlarged unit cell, which will have significantly reduced bond dimensions. We show particular examples in Fig.~\ref{fig:FigureCTMRG1}, where two different mappings are pointed out: the first one maps the structure to the single-layer tensor network with the $3 \times 3$ unit cell and the maximal scaling with the bond dimension as $\max(\chi_{b} D, D^{3})$, while the second one results in the $4\times4$ unit cell and the maximal scaling with the bond dimension as $\chi_{b} D$. This results in the reduced cost of the CTMRG calculation, which scales only as $\chi^{3} D^{3} \chi_{b}^{3} + \chi^{2} (D^{6} + D^{2} \chi_{b}^{4})$.  If the CTMRG bond dimension $\chi$ has the scaling $\chi \propto \chi_{b}^{2}$, while $\chi_{b} \propto D$, then the total cost scales as $D^{12}$, which is tractable up to sufficiently large values of $D$. 

\change{We should note that the introduction of enlarged unit cells leads to the constant overhead in the computational time and memory requirements (e.g., these increase in 16 times for the $4\times 4$ unit cell). Still, in practice, the differences in scalings with the bond dimension become more crucial than these constant factors. In addition, the developed approach has the full potential to be generalized to the iPEPS wave functions with the larger original unit cells. For example, the $4 \times 4$ mapping of the original unit cell of the size $l \times m$ (after projection on the plane) results in the single-layer tensor network with the unit cell of the size $4l \times 4 m$ (see Ref.~\cite{2019_Hagshenas} for 2d case). }

\change{The converged CTMRG environments can also be used to obtain the reduced density matrices of the 3d iPEPS wave functions, which can be further employed in the local observables computation. For this purpose, one should leave open physical indices of the bulk tensor~$T$ on one particular site and replace all other sites (and boundary iPEPS tensors) with the CTMRG environments. The separation of the tensors $T$ and $T^{\dagger}$ to different sites of the enlarged unit cell and the combination of physical indices with the virtual indices may lead to potential asymmetry between the wave-function tensors and their conjugates. This may become an additional source of the absence of hermiticity or positivity in the resulting reduced density matrices. Still, in practice, these non-Hermitian parts of the density matrices converge to zero with an increase of the CTMRG bond dimension~$\chi$.   }

\section{Results for the Heisenberg model}\label{sec:Results}
Here, let us briefly discuss the application of the developed approach to the spin-1/2 isotropic Heisenberg model on the cubic lattice with the Hamiltonian
\begin{equation}
    H = J \sum_{\langle ij \rangle} {\bf S}_i\cdot{\bf S}_j ,
\end{equation}
where we set $J = 1$ (antiferromagnetic coupling), ${\bf S}_i$ are the local spin operators with the conventional relations to the Pauli matrices ($S^{\alpha}=\sigma^\alpha/2$ with $\alpha=X,Y,Z$), and $\langle ij \rangle$ runs over nearest-neighbor pairs of sites $i$ and $j$. 
The ground state of the model corresponds to the gapless antiferromagnetic phase with a two-site unit cell (containing sites $a$ and $b$). Still, \change{it can be mapped} to the system with a single-site unit cell by using the unitary transformation $R_a$ on all the sites of one sublattice ($i\in a$).  We choose the wave function to be real, so that the magnetization lies in the $XZ$ plane, and then perform the unitary transformation $R_a = \exp(i\pi S^{Y}) = i\sigma^{Y}$, which rotates the magnetization on each site $i\in a$ by $\pi$ angle around the $Y$ axis 
\change{(this ansatz is an example of more general spiral PEPS approach \cite{hasik2023incommensurateordertranslationallyinvariant}, which is also capable of describing incommensurate phases)}.  
After this unitary transformation, the bulk iPEPS becomes completely homogeneous with identical tensors on all the sites. Note also that the wave function preserves lattice rotational and reflection symmetries. We have used the ITensors numerical package \cite{ITensors} in our calculations. 

We optimize the iPEPS wave function with the repeated application of PEPO, which approximates the operator $\exp(-Hdt)$ with $dt = 0.02$. For the approximation, we employ the 3d generalization of the $W^I$ method from Ref.~\cite{2015_zaletel}. The convergence is generally reached in several tens of PEPO applications and the optimization time is much smaller than the consequent calculation of observables. 

\begin{figure}
    \includegraphics[width= \linewidth]{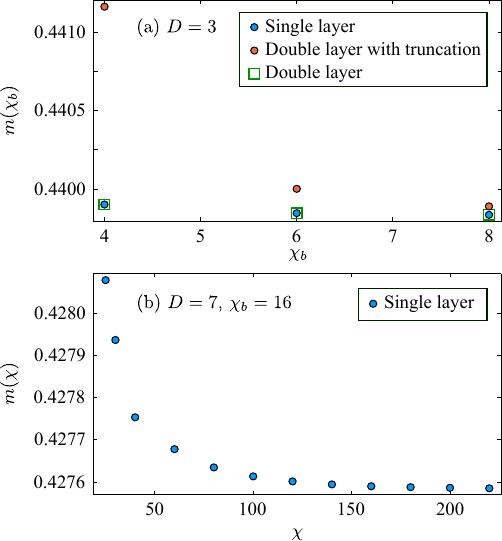} 
     \caption{\label{fig:Figure0}%
        (a) Convergence of the on-site magnetization with the boundary iPEPS bond dimension $\chi_{b}$ for the iPEPS wave function with $D=3$ and with different methods of CTMRG computation: single layer ($4\times4$ unit cell), double layer ($2\times2$ unit cell) and double layer with truncation ($2\times 2$ unit cell, but with bond dimensions of tensors truncated to $\chi_{b}$ with the simple update canonical form, as discussed in the text. (b) Convergence of the on-site magnetization with CTMRG bond dimension $\chi$ for the iPEPS wave function with $D=7$ and $\chi_{b} = 16$.}
\end{figure}

After obtaining the optimized iPEPS wave functions, we can compare different methods of calculating observables. To this end, let us focus on the iPEPS wave function with $D=3$, for which it is affordable to use different computational schemes. We first discuss the convergence of these results with $\chi_{b}$ and $\chi$ and also compare them with each other. 
In particular, in Fig.~\ref{fig:Figure0}(a) we show the results for several dimensions $\chi_{b}$ obtained with different methods of computation. The results are collected after reaching convergence in $\chi$. It is clear that the results of methods without truncation are in complete agreement, which proves the correct convergence of these methods. It should be noted that the double-layer construction is much more costly in terms of computational time than the single-layer method. Hence, in the computations below we always employ the single-layer method. The double layer with projection, in principle, has the same computational cost, as the single layer approach, but the necessity of truncation diminishes its accuracy.  Typically, one needs much higher $\chi_{b}$ for the double-layer method with truncation to reach approximately the same accuracy. In our calculations the dimension $\chi_{b}$ is the largest limitation (we could easily reach larger $D$ or $\chi$, if not the memory requirements for larger $\chi_{b}$). Hence, the double-layer with truncation is less effective than the single-layer approach.

Let us now focus on the single-layer calculations and determine the characteristic values of $\chi_{b}$ and $\chi$, which yield reliable results. In Fig.~\ref{fig:Figure0}(b) we show the convergence of magnetization with $\chi$ for $D=7$ and $ \chi_{b} = 16$ (these were the maximal $D$ and $\chi_{b}$ we were able to reach). We see that the convergence is reached at $\chi \approx 200$. In general, throughout our analysis, we observed an approximate relation $\chi \approx \chi_{b}^{2}$ for the final convergence in $\chi$.

\begin{figure}
    \includegraphics[width= \linewidth]{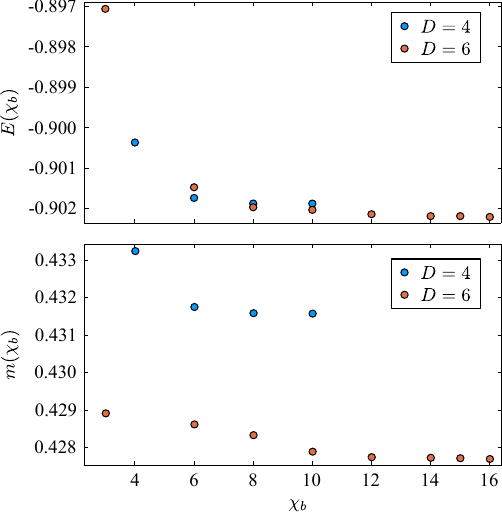} 
     \caption{\label{fig:Figure2}%
        Convergence of the energy per site $E$ and on-site magnetization~$m$ with the boundary iPEPS bond dimension $\chi_{b}$ for the iPEPS wave function of the bond dimensions $D=4$ and $D=6$. }
\end{figure}

In Fig.~\ref{fig:Figure2} we compare the convergences of the energy per site $E$ and local magnetization~$m$ with $\chi_{b}$ for two different bond dimensions~$D$. At $D=4$ the convergence is reached already at $\chi_{b} = 8$, while at $D=6$ it is approached at $\chi_{b} \approx 14-15$ (at $D=5$ it is $\chi_{b} = 12$). From here we conclude that the necessary condition for the convergence in $\chi_{b}$ can be expressed as $\chi_b\gtrsim 2 D$. Unfortunately, we \change{cannot} state this scaling precisely, furthermore, we expect that its determination will require a detailed study of convergence for other lattice models.

Our next goal is to collect the results for different bond dimensions~$D$ and to extrapolate them to the infinite-$D$ limit. The correlation length $\xi$ is typically viewed as a reliable measure to extrapolate the calculated observables. Note that in the single-layer approach, the correlation length can be obtained at the same computational cost as in 2d problems. 
\change{For this purpose, we employ the scaling relations from Ref.~\cite{Hasenfratz1993} (while the procedure for extrapolation of iPEPS results with the correlation-length scaling was proposed in Ref.~\cite{2018_rader})}:
\begin{eqnarray}
    E(\xi) = E(\infty) + a/\xi^{4},
    \\
    m^{2}(\xi) = m^{2}(\infty) + b/\xi^{2}.
\end{eqnarray}

\begin{figure}[t]
    \includegraphics[width= \linewidth]{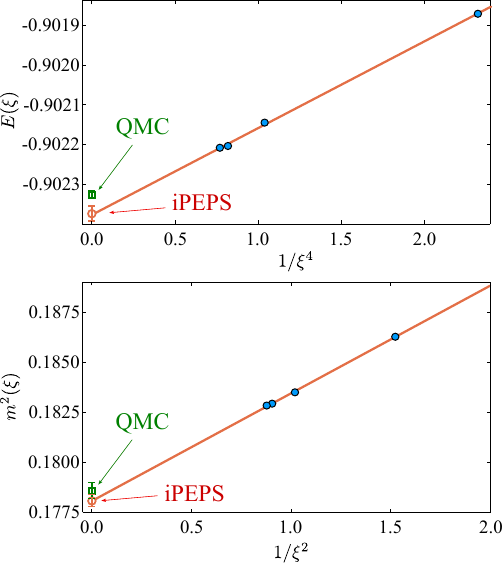} 
     \caption{\label{fig:Figure4}%
        Dependence of the energy $E(\xi)$ and square of magnetization $m^2(\xi)$ per site obtained with iPEPS for $D\in[4,7]$ (circles) with the corresponding linear fits to the powers of the correlation length $\xi^{-4}$ and $\xi^{-2}$, respectively. The estimated extrapolated values (open circles with error bars) are $E = - 0.90237(2)$ and $m^{2} = 0.1781(3)$. QMC data points (open squares with error bars) are taken from Ref.~\cite{2021_vlaar}.}
\end{figure}

The results of the fit are shown in Fig.~\ref{fig:Figure4}. In the limit $\xi\to\infty$ we obtain $E = -0.90237(2)$ for the energy and $m^{2} = 0.1781(3)$ for the magnetization per site. The fitting of the data is performed from the iPEPS simulations with $D \in[4,7]$ (\change{our results for $D=3$ with $\xi=0.53$, $E(\xi)=-0.89960$, and $m^2(\xi)= 0.1935$, which are obtained at $\chi_{b}=8$, appear not completely consistent with the fitting ansatz and not shown in Fig.~\ref{fig:Figure4} for the sake of visibility of the main data}). Let us additionally note that the correlation length is calculated for the largest available $\chi_{b}$ and $\chi$ at the given $D$ and is not additionally extrapolated to the limits $\chi\to\infty$ and $ \chi_{b}\to\infty$. Hence, it could explain possible minor errors in the estimated correlation length values. Nevertheless, our estimates agree \change{relatively} well with the available published results from the quantum Monte-Carlo approach, $E_{\rm QMC} = -0.902325(11)$ and $m^{2}_{\rm QMC} = 0.1786(4)$~\cite{2021_vlaar}. 
\change{One can also compare our results with the extrapolated iPEPS values from Ref.~\cite{2021_vlaar}: $m^{2} = 0.1826(2)$ and $E = - 0.9024(1)$. We see that the energy estimates agree well with those from Ref.~\cite{2021_vlaar}, while the magnetization is significantly lower indicating that the estimates are closer to the QMC prediction (in fact, at $D=6$ our data for $m^2$ practically approach the extrapolated iPEPS result from Ref.~\cite{2021_vlaar}). } 

\section{Conclusions}

In this study, we performed the tensor network calculations for the 3d Heisenberg model on the cubic lattice. We intensively used the multilayer structure of the tensor network for the calculation of observables in order to diminish the computational cost both during the boundary iPEPS search and during the final CTMRG contraction. As a result, we were able to perform calculations up to $D=7$ without the usage of U(1) symmetries. 

There are several research directions for the future studies. In particular, introducing U(1) symmetries \cite{Singh_2010, Singh_2011} may allow reaching even higher bond dimensions~$D$. Additional research directions concern the algorithm development for larger unit cells, as well as its application to other lattice geometries and fermionic models \cite{Corboz_fermions1, Corboz_fermions2, Corboz_fermions3}. Furthermore, it would be also interesting to apply the full update optimization~\cite{Jordan_2008, 2015_phien} of 3d tensor networks by means of the proposed single-layer CTMRG environments. The latter looks manageable to us, at least, for $D=3$ (and possibly for $D=4$). It may be also possible to generalize the multilayer approach of observables calculation to finite temperature quantum systems \cite{2020_jahromi, 2021_jahromi} and to classical 3d statistical mechanics models \cite{NISHINO2000, Nishino2001, Vanhecke_2018}. 
\change{Another relevant research direction is a more comprehensive analysis of the scalings of the CTMRG and boundary iPEPS bond dimensions with $D$, since the results reported in this study should be considered as particular observations valid for the specific 3d Heisenberg model.} 

\acknowledgements
The authors acknowledge support by the National Research Foundation of Ukraine under the call ``Excellent Science in Ukraine'', project No.~2023.03/0073 (2024--2026).

\appendix

\change{
\section{$W^{I}$ PEPO construction}\label{app:A}

Let us briefly discuss how to generalize the construction of the matrix product operator (MPO) of the type~$W^{I}$ for the evolution (time-stepper) operator $\exp(-H dt)$ \cite{2015_zaletel} to 3d PEPO. The main idea of the $W^{I}$ MPO approach is to construct an operator, which properly captures all terms in the expansion of $\exp(-H dt)$, which are of the first order in $dt$ and all terms of the higher orders in $dt$ with the condition that the support of individual operators in the expansion does not overlap. For example, let us take the Hamiltonian $H = \sum_{\langle ij \rangle} h_{ij}$, where the sum runs over all nearest neighbor pairs of sites. Then, this construction captures all terms of the type $dt^{2} h_{ij} h_{kl}$ if all the sites $i,j,k,l$ are different (i.e., the support of operators $h_{ij}$ and $h_{kl}$ does not overlap). Hence, the natural generalization of the $W^{I}$ method to PEPO on arbitrary lattices should also capture all these operators with non-overlapping support. 

We are interested in the $W^{I}$ construction for the Heisenberg model on the cubic lattice. First, note that, since our iPEPS is constructed by means of additional unitary transformations $i\sigma^{Y}$ on one sublattice (see Sec.~\ref{sec:Results}), the effective Hamiltonian on the bond with account of these rotations becomes $h_{kl} = -S_{k}^{X} S_{l}^{X} - S_{k}^{Z} S_{l}^{Z} - (i S_{k}^{Y} )(i S_{l}^{Y})$. In Fig.~\ref{fig:WI} we show the non-zero PEPO terms for the Heisenberg model (there we show only non-zero terms for the $x$ axis of the cubic lattice; the non-zero terms for other axes are constructed analogously). 
\begin{figure}[t]
    \includegraphics[width= \linewidth]{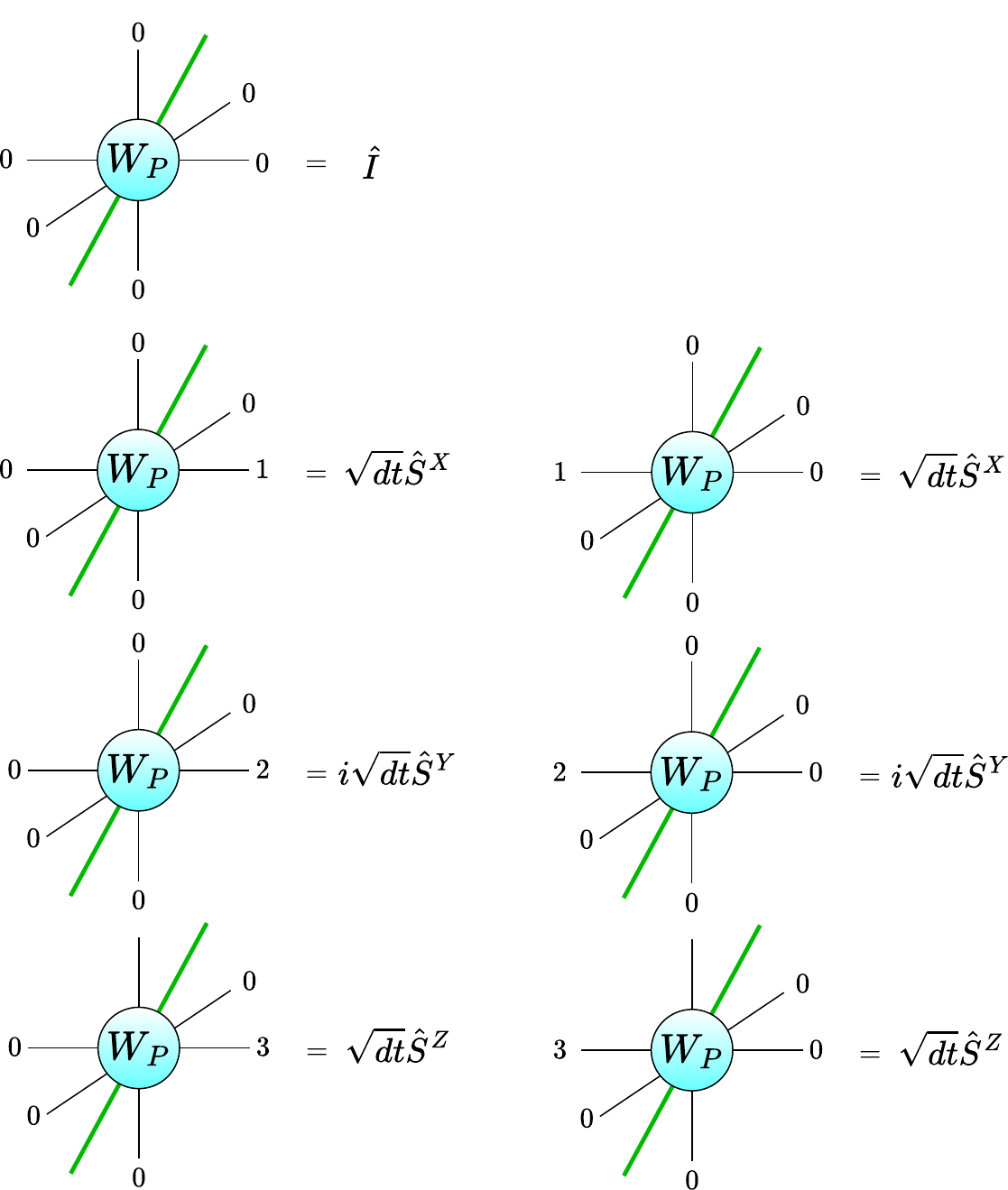}      \caption{\label{fig:WI}%
        Construction of the $W^{I}$ PEPO for the Heisenberg model. We fix virtual indices to the values $0,1,2,3$ and obtain operators acting on the physical indices. Note that here we show only terms corresponding to operators acting on the bonds along the $x$ axis of the cubic lattice. The similar terms corresponding to $y$ and $z$ axes have the same structure. All other terms (e.g., with several indices equal to $1$) vanish. }
\end{figure}

Within this approach, the PEPO has a virtual bond dimension $4$. If all virtual indices are set to zero, PEPO acts on the given site as the identity operator. Next, e.g., if the right virtual index is equal to one, then the PEPO site to the right must have its left index to be equal to one, which results in the operator $dt S_{i}^{X} S_{j}^{X}$ (if the index is equal to two or three, this results in $-dt S_{i}^{Y} S_{j}^{Y}$ or $dt S_{i}^{Z} S_{j}^{Z}$, respectively). Since all other virtual indices for these two sites are zero, the operators on other sites cannot overlap with the operator $S_{i}^{X} S_{j}^{X}$. However, all zero virtual indices also mean that the other sites are completely indifferent to the presence of the operator $ S_{i}^{X} S_{j}^{X}$. Hence, all types of operators $h_{ij} h_{kl}$ appear in the PEPO expansion as long as the sites $i,j,k,l$ are different. 

This PEPO construction has several important properties: it is real-valued and it is symmetric under reflections and rotations. Hence, the PEPO application to 3d iPEPS wave function does not spoil its properties and these can be preserved in the process of optimization. There are several possible generalizations of this construction to other systems. In particular, it can be modified to capture interactions beyond nearest neighbors or even exponentially decaying long-range interactions. Other methods from Ref.~\cite{2015_zaletel}, in particular, the $W^{II}$ approach can also be generalized to different lattices, but the precise recipe of this generalization is beyond the scope of the current study. 
}

\bibliography{3DPEPS}
\end{document}